\documentclass[12pt,fleqn]{article}
\usepackage{amsmath}
\usepackage{slashed}
\usepackage{url}
\usepackage{bm}
\usepackage{graphicx}
\usepackage[utf8]{inputenc}
\usepackage[english]{babel}
\usepackage{amsmath}
\usepackage{amsfonts}
\usepackage{amssymb}
\usepackage{graphicx}
\usepackage{float}
\usepackage{hyperref}
\usepackage{cite}
\allowdisplaybreaks

\begin{document}

\title{Local scale symmetry and black holes  }
\author{Malik Almatwi\footnote{malik.almatwi@gmail.com, Mobile: +963/0954335849\\ https://orcid.org/0000-0002-5403-683X}\,\,  }

\date{Department of Theoretical Physics, Faculty of Science, Damascus University, P.O. Box 30621, Damascus, Syria}
\maketitle

\tableofcontents

\section*{Abstract}
We derive solutions for a black hole coupled to a real scalar field under considerations of local scale symmetry. Starting with the simple case $(\lambda=0)$, we find the solutions and derive the black hole's temperature, entropy, and its effective mass, and observe an adiabatic accretion of the scalar field onto the black hole. As a result, the effective mass of the black hole grows. In the general case $(\lambda\ne 0)$, the situation becomes more intricate; The horizons (the vanishing points of $g_{tt}$) depend on the mass of the black hole, scalar field strength, $\lambda$ and other positive parameter $R_\phi$. In this case, we propose two possible behaviors: When the black hole's mass is small enough, the field is mostly concentrated far from the black hole, whereas when the black hole's mass is sufficiently large, it is mostly concentrated inside the black hole. 
\\

{\bf Key words:}  Local scale symmetry; Black hole; Scalar field; Thermodynamics; Effective mass.

\section{Introduction}
Global scale symmetry implies that a system's physical laws and dynamics remain the same when quantities such as length, time, and energy are multiplied by a constant factor. As a result, it preserves a system's behavior identically no matter what sizes and scales are used to take the measurements \cite{ r01, r02, r05, r00006}. However, this symmetry is not present in nature; it is instead thought to be a broken symmetry. In fact, this symmetry is valid in classical massless field theories (such as electromagnetic fields), in which the equations of motion in flat spacetime (Lorentz geometry) remain unchanged when the coordinates undergo a global scaling. This is theoretically in line with physics of standard model; The standard model's action is consistent with scale symmetry when the Higgs mass component is left out. This supports the idea that the mass component could originate from a vacuum expectation value of some scalar field $\phi$ within a fully scale-invariant proposal \cite{r11d, r11e, r0011f, r11f}.
\\

In the physics of black holes, conformal symmetry is essential, as it provides the mathematical basis for holographic dualities, with the AdS/CFT correspondence being the most prominent example. At the horizon, conformal symmetry appears, where the metric approximation naturally displays an Anti-de Sitter geometry ($AdS_2 \times S^2$). The boundary of Anti-de Sitter (AdS) space exhibits exact conformal symmetry. Serving as a hologram, the conformal field theory (CFT) on this boundary fully describes the quantum states of the black hole in the interior. Conformal symmetry in two dimensions (2D CFT) is exceptionally powerful, since its symmetry group is infinite. By means of the Cardy formula, physicists are able to count the microstates of a conformal field theory, discovering that they match exactly the Bekenstein-Hawking entropy ($S = A/4G$) of the corresponding black hole. This offers a concrete quantum explanation for the temperature and entropy of black holes  \cite{rh013, rh0013, rh13, rh14, rh15}.
\\

On the experimental side, a visible demonstration of scale invariance has recently appeared; Measurements of the Cosmic Microwave Background (CMB) \cite{ r11b, r11c} show that the power spectrum of the initial curvature perturbations is scale-invariant. This scale invariance serves as an essential benchmark for testing whether models of the early universe are correct. Among cosmological models of the early universe, scale invariance is put forward in the Starobinsky model, which ranks as one of the most successful inflationary models  \cite{rh1, rh2, rh3, rh4}. The Starobinsky model combines the Einstein-Hilbert action with a higher-order Ricci curvature term, namely the $R^2$ term. As a result, this model accounts for the accelerated expansion that took place during the early Universe. Since inflation of the early universe is thought to happen at high energy densities, the quadratic term $R^2$ (which supplies an invariant term) grows important, as it points to the presence of a gravitational scale-invariant theory. Furthermore, this symmetry has been shown to be capable of dynamical breaking, which produces General Relativity once inflation ends \cite{rh5, rh6}.
\\

In general relativity physics, spacetime is curved; thus, the local Lorentz symmetry is employed. Therefore, in order to satisfy the equivalence principle of gravity, the global scale symmetry needs to be generalized to a local scale symmetry, leading to a modification of Einstein gravity by including other symmetries rather than just invariance against the local transformations of the coordinates and diffeomorphisms. Including other interactions, such as the coupling of gravity with real scalar field, it is possible to construct local conformal invariance in the theory of gravity. Moreover, the coupling with a scalar field is widely used in cosmology in order to explain the inflation and accelerated expansion of the universe, and implementing conformal symmetry can act as an alternative framework to the standard dark energy by using geometric scaling rules to drive cosmic acceleration instead of an unseen repulsive force \cite{rh06, rh07}.
\\

Typically, to establish local scale invariance (local conformal or Weyl invariance), we need to get rid of dimensionful parameters (such as masses, Planck's constant, Newton's constant, etc.) by substituting them with a real scalar field. This creates a theory where the rules of physics remain unchanged under locally scaling the metric tensor (a spacetime-dependent transformation) \cite{r22}-\cite{rh21}. In local scale symmetry theories, it is possible, as we will observe in this study, to get solutions for a black hole by solving Einstein's equation in a static spherically symmetric system employing spherical coordinates and deriving the event horizons, with noticing that scaling the metric tensor with a non-zero everywhere differentiable function does not remove the event horizons.
\\

In certain scalar-tensor cosmological models where a scalar field is coupled to the Ricci scalar, Jackiw and Pi showed that local conformal invariance (Weyl invariance) is a "fake" symmetry because its associated Noether currents vanish identically, therefore this symmetry lacks physical consequences and has no true dynamical role in the equations of motion. It acts more like a field redefinition or artifact of the parametrization than a genuine physical gauge freedom \cite{r56}. In this paper, although this symmetry is also non-dynamical (its current vanishes), the solution we obtain could not be locally scaled to Schwarzschild (vacuum solutions) nor de Sitter solutions (cosmological constant solutions). Thus, the local scale symmetry in our solutions is not a trivial symmetry.
\\

A simple model for using the concept of local scale symmetry with a real scalar field on curved spacetime is constructed in the action \cite{r56}:
\begin{equation}\label{eq:53}
S = \int {{d^4}x\sqrt { - g} \left( {\frac{1 }{{12}}{\phi ^2}R + \frac{1}{2}{g^{\mu \nu }}{\partial _\mu }\phi {\partial _\nu }\phi  - \frac{\lambda }{8}{\phi ^4}} \right)} \, .
\end{equation}
Here $R$ denotes the Ricci curvature scalar, and $\phi$ is a real scalar field. This action remains invariant under ${g_{\mu \nu }}( x ) \to \Omega^2 ( x ){g_{\mu \nu }}( x )$ and $\phi( x ) \to \Omega^{-1} ( x ) \phi( x )$ up to boundary terms, for a nonzero everywhere differentiable function $\Omega ( x )$. We regard this transformation as an internal symmetry performed in the space $({g_{\mu \nu }}( x ), \phi( x ))$ over the same coordinates $x$, rather than a coordinates transformation.
\\

In Ref.\cite{r56}, it was shown that the Noether current linked with scale invariance of the action (\ref{eq:53}) vanishes, indicating that the associated symmetry lacks any dynamical aspect, and the Einstein equation of this action is consistent with the equation of motion of the field $\phi$. Therefore, it suffices to solve the Einstein equation. The Einstein equation of the action (\ref{eq:53}) is
\begin{equation}\label{eq:63}
\frac{1 }{6}{\phi ^2}{G_{\mu \nu }} + \frac{{2 }}{3}{\partial _\mu }\phi {\partial _\nu }\phi  - \frac{{1 }}{6}{g_{\mu \nu }}{\left( {\partial \phi } \right)^2} - \frac{1 }{3}\phi \left( {{\nabla _\mu }{\nabla _\nu } - {g_{\mu \nu }}{\nabla ^2}} \right)\phi  +\frac{\lambda}{8} \phi^4 g_{\mu \nu }=  0 \, ,
\end{equation}
\\
where $( {\partial \phi } )^2=g^{\mu \nu } \partial _\mu \phi \partial _\nu \phi$ and $\nabla ^2  = g^{\mu \nu } \nabla _\mu  \nabla _\nu  $. The connection $\nabla$ associates solely with the employed metric $g_{\mu \nu }$ through $\nabla _\mu g_{ \nu \rho} =0$.
\\

The equation (\ref{eq:63}) has the solution 
\begin{equation}
G_{\mu \nu }  + {{3\lambda }}\phi _0^2 g_{\mu \nu } /4 = 0$ \, , \,\, \, $\phi=\phi_0=\text{constant}\ne 0\, , 
\end{equation}
where $\nabla _\mu \phi_0 =\partial _\mu \phi_0=0$. But the equation $G_{\mu \nu }  + {{3\lambda }}\phi _0^2 g_{\mu \nu } /4 = 0$ possesses a Schwarzschild-de Sitter solution in a spherically symmetric static system\footnote{When $\lambda=0$, the equation $G_{\mu \nu } = 0$ (so $R = 0$) is not a vacuum equation for the action (\ref{eq:53}); this action does not vanish in this case because of the boundary terms that arise from the local scale transformations.}. Thus, it becomes clear that the action (\ref{eq:53}) extends the Schwarzschild-de Sitter solution, where the constant energy density ${{3\lambda }}\phi _0^2  /4$ becomes a space-time dependent energy density, and the global scale symmetry of the Schwarzschild-de Sitter solution extends to local scale symmetry. This is evident in the gauged solution of the Schwarzschild-de Sitter solution:
\begin{equation}
ds^2  =  - \Omega ^2 \left( r \right)\left( {1 - \frac{{2M}}{r} - \frac{{\phi _0^2 \lambda }}{4}r^2 } \right)dt^2  + \frac{{\Omega ^2 \left( r \right)}}{{1 - \frac{{2M}}{r} - \frac{{\phi _0^2 \lambda }}{4}r^2 }}dr^2  + \Omega ^2 \left( r \right)r^2 d\Omega _2^2 \, ,
\end{equation}
with
\begin{equation}\label{eq:118}
\phi \left( r \right) = \frac{{\phi _0 }}{{\Omega \left( r \right)}}\, .
\end{equation}
This solution satisfies Einstein's equation (\ref{eq:63}) for any non-zero everywhere differentiable function $\Omega ( r )$. This defines the pure gauged solution (\ref{eq:118}) as a non-zero everywhere differentiable function. We attribute this feature to the local scale symmetry, which allows for the free redistribution of the field $\phi$ (and thus its energy) throughout space. In other words, with local scale symmetry, the scalar field is not required to concentrate near any certain points in space (no preferred locations for $\phi$ to concentrate within).
\\

Because the local scale transformations in (\ref{eq:118}) are performed with fixing the mass $M$ and the value $\phi _0$, thus it is an internal symmetry performed in the space $({g_{\mu \nu }}, \phi)$ in the same coordinates $t, r,...$\, . This means that the transformation ${g_{\mu \nu }}( x ) \to \Omega^2 ( x ){g_{\mu \nu }}( x )$, $\phi( x ) \to \Omega^{-1} ( x ) \phi( x )$ changes the strength of the gravity with changing the energy of the field $\phi$ and there is no need to transform the coordinates. 
\\

\textit{Remark}: Regarding the pure gauged solution (\ref{eq:118}); If a solution $\phi(r)$ vanishes or is undefined at some positions in space, it could not gauge to Schwarzschild nor cosmological constant solutions. In fact, in this study, the obtained solution of the field $\phi(r)$ is of the form
\begin{equation}\label{eq:114}
\phi(r)=h^{-1}/(r-R_\phi)\, ,
\end{equation}
for some arbitrary parameters $h\ne 0$ and $R_\phi>0$. The solution (\ref{eq:114}) is undefined on a sphere of radius $r = R_{\phi}$ and heavily concentrated near it, so it cannot be locally scaled to Schwarzschild (vacuum solutions) or to de Sitter solutions (cosmological constant solutions). This arises from the fact that scaling the solution (\ref{eq:114}) by any non-zero everywhere differentiable function $\Omega ( x )$ does not eliminate the divergence of the field (\ref{eq:114}) at the point $r=R_\phi$. Conversely and for the same reason, we cannot obtain the solution (\ref{eq:114}) from Schwarzschild or de Sitter solutions.
\\

This results in a possibility of obtaining a black hole phenomenon in local scale symmetry theories, noting that scaling the metric tensor by a non-zero everywhere differentiable function does not eliminate event horizons. Furthermore, in black hole physics, it is meaningful to test the global scale transformation; $\Omega ( x )=\alpha=\text{constant}$. For instance, under a global scale transformation $t\to \alpha t$ and $r\to \alpha r$, the parameter $R_\phi$ scales to $\alpha R_\phi$, and the black hole's mass $M$ similarly scales to $\alpha M$. This shifts the location of the main concentration of the scalar field $\phi$, thus this transformation shifts the distribution of the field $\phi$ throughout space. Meanwhile, the conditions $R_\phi<2M$ or $R_\phi>2M$ (or $R_\phi>>2M$) remain unchanged under this scaling.
\\

A notable aspect of the solution (\ref{eq:114}) is that the parameter $R_\phi >0$ and the black hole's mass $M$ are independent of each other, and values of $R_\phi$ are not bound to any range. Therefore, there are no restrictions on the field $\phi$ to accumulate at any location in space. We link this property to the local scale invariance, which implies that the black hole's energy (given by its mass) and the energy of the field $\phi$ (which depends on $R_\phi$) are independent of one another. This resembles the Schwarzschild-de Sitter solution, where the constant energy density ${{3\lambda }}\phi _0^2  /4$ is independent of the black hole's mass $M$.
\\

To ensure that our results describe the black hole, we derive the speed and force of an object in free fall (the derivation is after equation (\ref{eq:62})), we obtain (for $\lambda=0$)
\begin{equation}
\left( {\frac{{dr}}{{d\tau }}} \right)^2  = \frac{1}{{\left( {1 - \kappa } \right)^2 }}E^2  - \frac{1}{{\left( {1 - \kappa } \right)}}\left( {1 - \frac{r}{{R_\phi  }}} \right)^2 \left( {1 - \frac{{2M}}{r}} \right) \, ,
\end{equation}
and
\begin{equation}\label{eq:115}
\frac{{d^2 r}}{{d\tau ^2 }} + \frac{1}{{\left( {1 - \kappa } \right)}}\frac{1}{r}\left( {\frac{M}{r} + \frac{M}{R_\phi} - \frac{r}{R_\phi}} \right)\left( {1 - \frac{r}{{R_\phi  }}} \right) = 0 \, ,
\end{equation}
where $\kappa= 2M/R_{\phi} $. We see that when $R_{\phi}>>2M$ where the field $\phi$ resides far away from the black hole, the velocity and force of an object in free fall reduce to those of Schwarzschild solution. On the other hand, when $R_{\phi}\approx 2M$, the field $\phi$ is found mostly near the black hole, and the free falling force in $r> R_{\phi}>2M$ always points toward $r=0$, so our solution describes a black hole whose center is $r=0$.
\\

The singularities of the scalar field identified in our analysis align with those predicted by the Brans-Dicke scalar-tensor theory which indicates that when a spacetime metric is cast into a Class 1 embedding condition (or represents Class 1 solutions such as in anisotropic or spherically symmetric compact star/collapse models), the scalar field $\phi$ and the spacetime geometry can feature a spatial dependence that diverges, vanishes, or hits a zero/singularity at the central origin ($r\to 0$ or at a specific boundary/throat radius) \cite{r58, r59, r60}.
\\

We notice some differences between the cases $\lambda=0$ and $\lambda\ne0$. When $\lambda=0$, Einstein's equation (\ref{eq:63}) becomes symmetric under scaling the field $\phi$ by an arbitrary non-zero constant. While when $\lambda\ne0$, it becomes asymmetric under that scaling; consequently, the solutions for $\lambda\ne0$ depend on the strength of the field $\phi$. As we will see, when $\lambda\ne0$, the strength of the field $\phi$ determines whether $\phi$ tends to concentrate near the black hole or away from it.

\section{Solution of black hole of local scale symmetry of $\lambda= 0$} 
In the previous section, we demonstrated the possibility of obtaining black hole phenomena in local scale symmetry theories with noting that scaling the metric tensor with a non-zero everywhere differentiable function does not eliminate event horizons. And if a solution $\phi(r)$ vanishes or is undefined at some points in space, the metric solution cannot be scaled to Schwarzschild solution. In this study, the solution we derive for $\phi$ is undefined on a surface of radius $r = R_{\phi}$ and is unaffected by the black hole's horizon, as it is a continuous differentiable function at $r=2M$. We attribute this feature to local scale symmetry, through which, as described below equation (\ref{eq:118}), the field $\phi$ is not restricted to concentrate near any certain points in space (no preferred locations for $\phi$ to concentrate within). We find that the static field $\phi$ possesses its own horizon, where $\phi^{-1}$ becomes zero. In solutions with $\lambda=0$, we determine that the location of $\phi^{-1}=0$ coincides with that of $g_{tt}=0$. While for $\lambda\ne 0$, it does not.
\\

In this section, we obtain solutions for the static scalar field $\phi$ and the black hole (so static metric), for $\lambda=0$, in spherical symmetric system. We then derive the free-fall acceleration and observe that the scalar field behaves as if it follows that force.
\\

We begin with the Einstein's equation (\ref{eq:63}):
\begin{equation}\label{eq:90}
\frac{1 }{6}{\phi ^2}{G_{\mu \nu }} + \frac{{2 }}{3}{\partial _\mu }\phi {\partial _\nu }\phi  - \frac{{1 }}{6}{g_{\mu \nu }}{\left( {\partial \phi } \right)^2} - \frac{1 }{3}\phi \left( {{\nabla _\mu }{\nabla _\nu } - {g_{\mu \nu }}{\nabla ^2}} \right)\phi  +\frac{\lambda }{8}{\phi ^4}{g_{\mu \nu }}= 0\, .
\end{equation}
We solve this equation in a static spherically symmetric system, thereby obtaining black hole solutions with $r=0$ as its center, along with a scalar field horizon. In this case, all functions depend only on the radial coordinate $r$. This permits us to use the metric.
\begin{equation}\label{eq:1}
ds^2  = - u (r) d{t^2} +v(r) d{r^2} +  {r^2}d\Omega _2^2\, ,
\end{equation}
for some two independent functions $u(r)$ and $v(r)$. The Einstein's equation (\ref{eq:90}) gives (for the components $G_{tt }$, $G_{rr}$ and $G_{\theta \theta }$)
\begin{equation}\label{eq:93}
\begin{split}
&\frac{1}{6}\phi ^2 \left( r \right)G_{tt}  - \frac{1}{6}g_{tt} g^{rr} \left( {\frac{d}{{dr}}\phi \left( r \right)} \right)^2  - \frac{1}{3}\phi \left( r \right)\left( {\nabla _t \nabla _t  - g_{tt} \nabla ^2 } \right)\phi \left( r \right) +\frac{\lambda }{8}{\phi ^4}{g_{tt }}= 0\,,\\
&\frac{1}{6}\phi ^2 \left( r \right)G_{rr}  + \frac{2}{3}\left( {\frac{d}{{dr}}\phi \left( r \right)} \right)^2  - \frac{1}{6}g_{rr} g^{rr} \left( {\frac{d}{{dr}}\phi \left( r \right)} \right)^2 \\
&\quad\quad \quad\quad\quad- \frac{1}{3}\phi \left( r \right)\left( {\nabla _r \nabla _r  - g_{rr} \nabla ^2 } \right)\phi \left( r \right) +\frac{\lambda }{8}{\phi ^4}{g_{rr }}= 0\,,\\
&\frac{1}{6}\phi ^2 \left( r \right)G_{\theta \theta }  - \frac{1}{6}g_{\theta \theta } g^{rr} \left( {\frac{d}{{dr}}\phi \left( r \right)} \right)^2  - \frac{1}{3}\phi \left( r \right)\left( {\nabla _\theta  \nabla _\theta   - g_{\theta \theta } \nabla ^2 } \right)\phi \left( r \right)+\frac{\lambda }{8}{\phi ^4}{g_{\theta \theta }} = 0\,.\\
\end{split}
\end{equation}
The corresponding Einstein tensor $G_{\mu \nu}$ (of the metric (\ref{eq:1})) is
\begin{equation}\label{eq:91}
\begin{split}
&G_{tt}  = \frac{{r\frac{{\rm{d}}}{{{\rm{d}}r}}v\left( r \right) + v^2 \left( r \right) - v\left( r \right)}}{{r^2 v^2 \left( r \right)}}u\left( r \right)\,,\,\,\,G_{rr}  = \frac{{r\frac{{\rm{d}}}{{{\rm{d}}r}}u\left( r \right) - u\left( r \right)v\left( r \right) + u\left( r \right)}}{{r^2 u\left( r \right)}}\,,\\
&G_{\theta \theta }  = \frac{{2r^2 u\left( r \right)v\left( r \right)\,\frac{{{\rm{d}}^2 }}{{{\rm{d}}r^2 }}u\left( r \right) - r^2 v\left( r \right)\left( {\frac{{\rm{d}}}{{{\rm{d}}r}}u\left( r \right)} \right)^2  - r^2 u\left( r \right)\left( {\frac{{\rm{d}}}{{{\rm{d}}r}}u\left( r \right)} \right)\left( {\frac{{\rm{d}}}{{{\rm{d}}r}}v\left( r \right)} \right)}}{{4u^2 \left( r \right)v^2 \left( r \right)}}
 &\\
&\quad\quad\quad+ \frac{{2ru\left( r \right)v\left( r \right)\,\frac{{\rm{d}}}{{{\rm{d}}r}}u\left( r \right) - 2ru^2 \left( r \right)\,\frac{{\rm{d}}}{{{\rm{d}}r}}v\left( r \right)}}{{4u^2 \left( r \right)v^2 \left( r \right)}}\,.
 \end{split}
\end{equation}
We have
\begin{equation}
\nabla ^2 \phi \left( r \right) = g^{tt} \nabla _t \nabla _t \phi \left( r \right) + g^{rr} \nabla _r \nabla _r \phi \left( r \right) + g^{\theta \theta } \nabla _\theta  \nabla _\theta  \phi \left( r \right) + g^{\phi \phi } \nabla _\phi  \nabla _\phi  \phi \left( r \right)\, ,
\end{equation}
with
\begin{equation}\label{eq:92}
\begin{split}
& \nabla _t \nabla _t \phi \left( r \right) =  - \Gamma _{tt}^r \phi ' =  - \frac{1}{2}\frac{{u'}}{v}\phi '\, , \, \,\,\, \nabla _r \nabla _r \phi \left( r \right) = \phi '' - \Gamma _{rr}^r \phi ' = \phi '' - \frac{1}{2}\frac{{v'}}{v}\phi ' \, ,\\ 
& \nabla _\theta  \nabla _\theta  \phi \left( r \right) =  - \Gamma _{\theta \theta }^r \phi ' = \frac{r}{v}\phi '\,, \,\, \,\,\nabla _\phi  \nabla _\phi  \phi \left( r \right) =  - \Gamma _{\phi \phi }^r \phi ' = \frac{{r\sin ^2 \left( \theta  \right)}}{v}\phi '\, , \\ 
 \end{split}
\end{equation}
it yields
\begin{equation}
\nabla ^2 \phi \left( r \right) = \frac{{\phi ''}}{v} + \frac{1}{2}\left( {\frac{{u'}}{{uv}} - \frac{{v'}}{{v^2 }}} \right)\phi ' + \frac{{2\phi '}}{{rv}} \, .
\end{equation}
We have used the notation $F'=dF/dr$. By that, the Einstein's equation (\ref{eq:93}) becomes
\begin{equation}\label{eq:98}
\begin{split}
&\frac{{uv^2  + ruv' - vu}}{{6r^2 v^2 }}\phi ^2  + \frac{1}{6}\frac{u}{v}\phi '^2  - \frac{1}{3}\phi \left( {u\frac{{\phi ''}}{v} - \frac{{uv'}}{{2v^2 }}\phi ' + \frac{{2u\phi '}}{{rv}}} \right) - \frac{\lambda }{8}u\phi ^4  = 0\,,\\
 &\\
&\frac{{ru' - uv + u}}{{6r^2 u}}\phi ^2  + \frac{1}{2}\phi '^2  - \frac{1}{3}\phi \left( { - \frac{{u'}}{{2u}} - \frac{2}{r}} \right)\phi ' + \frac{\lambda }{8}v\phi ^4  = 0\,, \\
 &\\
& \frac{1}{6} \frac{{2r^2 uvu'' - r^2 vu'^2  - r^2 uu'v' + 2ruvu' - 2r\,u^2 v'}}{{4u^2 v^2 }} \phi ^2 \\ 
&\quad - \frac{{r^2 }}{{6v}}\phi '^2  + \frac{{r^2 }}{{3v}}\phi \phi '' + \left( {\frac{{r^2 }}{2}\frac{{u'}}{{uv}} - \frac{{r^2 }}{2}\frac{{v'}}{{v^2 }} + \frac{r}{v}} \right)\frac{{\phi \phi '}}{3} + \frac{\lambda }{8}r^2 \phi ^4  = 0\,.  
 \end{split}
\end{equation}
When $\lambda=0$, the equations (\ref{eq:98}) have the solution (See Appendix \ref{appendix:a} for details)
\begin{equation}\label{eq:33}
\begin{split}
&ds^2  =  - \frac{{\left( {1 - \frac{r}{{R_\phi  }}} \right)^2 \left( {1 - \frac{{R_{bh} }}{r}} \right)}}{{\left( {1 - \kappa } \right)}}dt^2  + \frac{{\left( {1 - \kappa } \right)}}{{\left( {1 - \frac{r}{{R_\phi  }}} \right)^2 \left( {1 - \frac{{R_{bh} }}{r}} \right)}}dr^2  + r^2 d\Omega _2^2 \,, \\
& \phi \left( r \right) = \frac{1}{{h\left( {r - R_{\phi}} \right)}}\, .
\end{split}
\end{equation}
For some arbitrary parameters $R_\phi >0$, $R_{bh}>0$ and $h\ne 0$. With $\kappa= R_{bh}/R_{\phi} <1$. Regarding the units, we have $[\phi]=[r^{-1}]$, $[R_\phi]=[R_{bh}]=[r]$ and $[h]=[\lambda]=0$. The parameter $h^{-1}$ indicates to strength of $\phi$.
\\

It is evident that when $R_\phi $ grows sufficiently large ($R_\phi \to \infty$, $\kappa= R_{bh}/R_{\phi} \to 0$), the metric solution (\ref{eq:33}) tends to Schwarzschild metric. This leads us to identify $R_{bh}$ with $2M$, where $M$ denotes the mass of the black hole. Thus, we obtain
\begin{equation}\label{eq:34}
ds^2  =  - \frac{1}{{\left( {1 - \kappa } \right)}}\left( {1 - \frac{r}{{R_\phi  }}} \right)^2 \left( {1 - \frac{{2M}}{r}} \right)dt^2  + \frac{{\left( {1 - \kappa } \right)}}{{\left( {1 - \frac{r}{{R_\phi  }}} \right)^2 \left( {1 - \frac{{2M}}{r}} \right)}}dr^2  + r^2 d\Omega _2^2 \,.
\end{equation}
This metric characterizes a black hole with mass $M$ that is coupled to a real scalar field, taking into account local scale symmetry, in the scenario where $\lambda = 0$. These results are consistent with the explanation given below equation (\ref{eq:118}); $R_\phi$ does not relate with $R_{bh}=2M$, and the concentration of the field $\phi$ (and therefore its energy density) in the space is neither confined to particular locations nor equivalent to Schwarzschild solution. The  metric (\ref{eq:34}) does not depend on the parameter $ h$ (which indicates to strength of $\phi $) because of the symmetry of Einstein's equation (\ref{eq:63}) (when $\lambda = 0$) under scaling the field $\phi $ by any non-zero constant. The condition $\kappa= 2M/R_{\phi} <1$ means that the field $\phi$ (when $\lambda = 0$) could not accumulate mostly inside the black hole. 
\\

The vanishing points of $ g_{tt}$ (namely $r= 2M$ and $r=R_ \phi$) are independent of one another, and $ g_{tt}$ changes sign at only one of these points, specifically at $r= 2M$. By contrast, when $\lambda\ne 0$, the situation is different: $ g_{tt}=0$ identifies three points, and these points are specified by $M$, $R_ \phi$, $h$, and $ \lambda$, with $ g_{tt}$ reversing its sign at each of them.
\\

\begin{figure}[H]
  \includegraphics[width=0.5\textwidth]{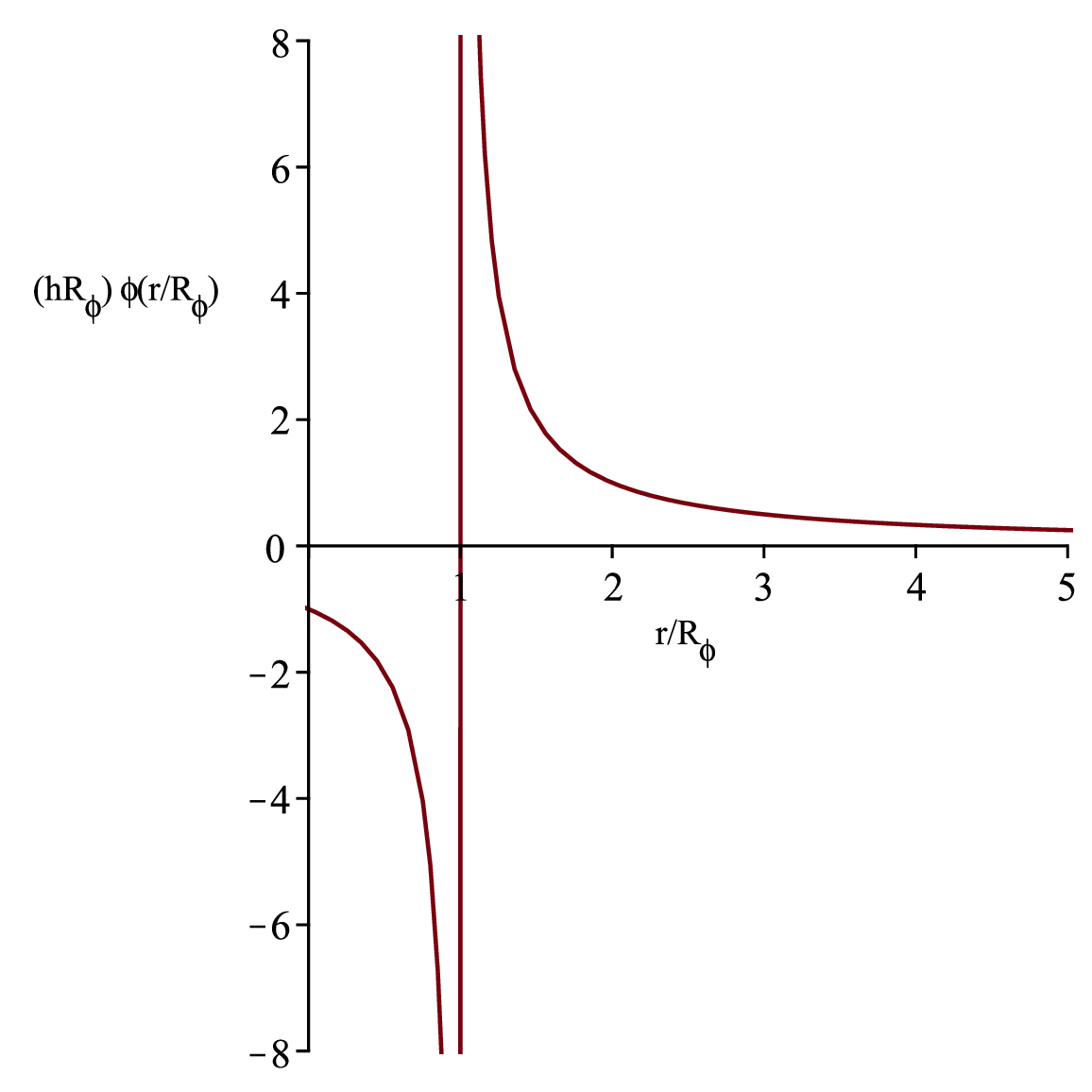}
  \caption{Behaviour of $\phi$ as a function of $r/R_\phi $.}
\label{fig:1}
\end{figure}

The behaviour of the field $\phi(r) $ is illustrated in figure (\ref{fig:1}); It remains undefined at the surface of radius $r=R_\phi $ and is concentrated in its vicinity. This behavior has nothing to do with the black hole; for $M=0$, we see the same behavior of $\phi(r) $. We can attribute this behavior to the self-gravitational force of $\phi(r) $; since it is an uncharged field, nothing stops it from gathering near any location in space because of its self-gravitational attraction. Negative values of $\phi$ pose no problem, because the action (\ref{eq:53}) is symmetric under $\phi \to -\phi$.

To figure out more about the behaviour of $\phi(r) $, let us test the force of free fall. Using the proper time $d\tau^2=-ds^2$, we get (for $d\Omega _2=0$)
\begin{equation}\label{eq:62}
 - 1 =  - \frac{1}{{\left( {1 - \kappa } \right)}}\left( {1 - \frac{r}{{R_\phi  }}} \right)^2 \left( {1 - \frac{{2M}}{r}} \right)\left( {\frac{{dt}}{{d\tau }}} \right)^2  + \frac{{\left( {1 - \kappa } \right)}}{{\left( {1 - \frac{r}{{R_\phi  }}} \right)^2 \left( {1 - \frac{{2M}}{r}} \right)}}\left( {\frac{{dr}}{{d\tau }}} \right)^2   \,.
\end{equation}
The vector ${\partial }/{{\partial t}}$ is a Killing vector, therefore $g_{tt} dt/d\tau$ is a conserved quantity, which implies
\begin{equation}\label{eq:063}
\left( {1 - \frac{r}{{R_\phi  }}} \right)^2 \left( {1 - \frac{{2M}}{r}} \right)\frac{{dt}}{{d\tau }} = E=\text{constant}\,.
\end{equation}
Using this formula in equation (\ref{eq:62}), yields
\begin{equation}\label{eq:064}
\left( {\frac{{dr}}{{d\tau }}} \right)^2  = \frac{1}{{\left( {1 - \kappa } \right)^2 }}E^2  - \frac{1}{{\left( {1 - \kappa } \right)}}\left( {1 - \frac{r}{{R_\phi  }}} \right)^2 \left( {1 - \frac{{2M}}{r}} \right) \, .
\end{equation}
Free fall's motion is characterized by the equation $\nabla_\tau (dx^\mu /d \tau)=0$. For $x^\mu=r$, this yields
\begin{equation}
\frac{{d^2 r}}{{d\tau ^2 }} + \Gamma _{rr}^r \left( {\frac{{dr}}{{d\tau }}} \right)^2  + \Gamma _{00}^r \left( {\frac{{dt}}{{d\tau }}} \right)^2  = 0\, .
\end{equation}
Using the equations (\ref{eq:063}) and (\ref{eq:064}) in this equation, we obtain
\begin{equation}
\frac{{d^2 r}}{{d\tau ^2 }} + \Gamma _{rr}^r \frac{1}{{\left( {1 - \kappa } \right)^2 }}E^2  + \Gamma _{00}^r E^2 \left( {1 - \frac{r}{{R_\phi  }}} \right)^{ - 4} \left( {1 - \frac{{2M}}{r}} \right)^{ - 2}  = 0 \, .
\end{equation}
We get the connections $\Gamma _{00}^r$ and $\Gamma _{rr}^r$ from the metric (\ref{eq:34}), they are
\begin{equation}
\Gamma _{00}^r  = \frac{1}{{\left( {1 - \kappa } \right)^2 r}}\left( {\frac{M}{r} + \frac{M}{R_\phi} - \frac{r}{R_\phi}} \right)\left( {1 - \frac{r}{R_\phi}} \right)^3 \left( {1 -\frac{{2M}}{r}} \right)\, ,
\end{equation}
and
\begin{equation}
\Gamma _{rr}^r  =  - \frac{1}{r}\left( {\frac{M}{r} + \frac{M}{R_\phi} - \frac{r}{R_\phi}} \right)\left( {1 - \frac{r}{R_\phi}} \right)^{ - 1} \left( {1 - \frac{{2M}}{r}} \right)^{ - 1} \, .
\end{equation}
Finally, we obtain
\begin{equation}\label{eq:67}
\frac{{d^2 r}}{{d\tau ^2 }} + \frac{1}{{\left( {1 - \kappa } \right)}}\frac{1}{r}\left( {\frac{M}{r} + \frac{M}{R_\phi} - \frac{r}{R_\phi}} \right)\left( {1 - \frac{r}{{R_\phi  }}} \right) = 0 \, .
\end{equation}
This equation governs the force experienced during free fall. To begin, let us consider the case where $M=0$ (meaning no black hole exists); The acceleration ${{d^2 r}}/{{d\tau ^2 }}$ is negative in $r>R_\phi$, but it becomes positive in $r<R_\phi$. This indicates an attractive force pointing toward the surface $r=R_\phi$; this force draws the field $\phi$ close to that surface, while leaving it undefined on the surface itself.
\\

When $M\ne 0$, the acceleration ${{d^2 r}}/{{d\tau ^2 }}$ determines the magnitude and direction of the force produced by the black hole and the scalar field. In $r>R_\phi$, this force is attractive (${{d^2 r}}/{{d\tau ^2 }}$ is negative), drawing the particles in from $r>R_\phi$ and keeping them near the surface $r=R_\phi$ on the outside. As a result, the field $\phi$, equation (\ref{eq:33}), is mostly concentrated around the surface $r=R_\phi$, both inside and outside it. This force becomes positive (${{d^2 r}}/{{d\tau ^2 }}>0$) in $r_0<r<R_\phi$, where $r_0$ is
\begin{equation}\label{eq:66}
r_0  = \frac{M}{2} + \frac{M}{2}\,\sqrt {1 + \frac{{4R_\phi  }}{M}}\, . 
\end{equation}
Therefore, it displaces the particles from $r_0<r<R_\phi$ and places them near the surface $r=R_\phi$ within. Finally, in the region $r<r_0$, this force is attractive (${{d^2 r}}/{{d\tau ^2 }}$ is negative), so it draws the particles toward the interior of the black hole, which explains why the field $\phi$, equation (\ref{eq:33}), does not vanish inside the black hole, specifically $\phi(0)=-1/(h R_\phi)$. In addition, at $r=r_0$, this force vanishes; a balance occurs between two opposing forces (one from the black hole and the other from the field $\phi$). We conclude that the behaviour of the coupled scalar field resembles that of a field obeying the free fall equation $D_\tau (dx^\mu /d \tau)=0$.
\\

For $R_\phi>>2M$ (when $R_\phi$ is in cosmological scales), the acceleration (\ref{eq:67}) in $2M<r<<R_\phi$ approximates to 
\begin{equation}
\frac{{d^2 r}}{{d\tau ^2 }} \approx  - \frac{1}{{\left( {1 - \kappa } \right)}}\frac{M}{{r^2 }}\, .
\end{equation}
So we define the effective mass of the black hole by
\begin{equation}\label{eq:119}
M_{eff}  = \frac{M}{{\left( {1 - \kappa } \right)}} = M\left( {1 - \frac{{2M}}{{R_\phi  }}} \right)^{ - 1} \, .
\end{equation}
In the section on thermodynamics, we come to the same definition by comparing the black hole's derived temperature with the standard Hawking's black hole temperature formula.

\section{Free fall force in local scale symmetry} 
In order to obtain free fall force with respecting the local scale symmetry, we use the action
\begin{equation}
S[\gamma ] = \frac{1}{2}\int\limits_\gamma  {\phi ^2 \left( x \right)g_{\mu \nu } \left( x \right)\dot x^\mu  \dot x^\nu  d\tau } \,,
\end{equation}
or
\begin{equation}\label{eq:1117}
S[x\left( \tau  \right) ] = \frac{1}{2}\int {\phi ^2 \left( {x\left( \tau  \right)} \right)g_{\mu \nu } \left( {x\left( \tau  \right)} \right)\dot x^\mu  \left( \tau  \right)\dot x^\nu  \left( \tau  \right)d\tau } \,,
\end{equation}
which is invariant under local scale transformations ${g_{\mu \nu }}( x ) \to \Omega^2 ( x ){g_{\mu \nu }}( x )$ and $\phi( x ) \to \Omega^{-1} ( x ) \phi( x )$ with keeping the path $\gamma$ unchanged. This action gives the free fall force
\begin{equation}\label{eq:1116}
\begin{split}
\nabla _\tau  \dot x^\mu  \left( \tau  \right) =  - 2\dot x^\mu  \left( \tau  \right)\frac{d}{{d\tau }}\ln \phi \left( {x\left( \tau  \right)} \right)\, + \left. {\left( {g^{\mu \nu } \left( x \right)\partial _\nu  \ln \phi \left( x \right)} \right)} \right|_{x\left( \tau  \right)}  \left( {g_{\rho \sigma } \dot x^\rho  \left( \tau  \right)\dot x^\sigma  \left( \tau  \right)} \right) \, ,
\end{split}
\end{equation}
where $ \dot x^\mu   \left( \tau  \right)=d x^\mu   \left( \tau  \right) /d\tau $. And $\tau$ is the proper time defined in $d\tau ^2  =  - \phi ^2 g_{\mu \nu } dx^\mu  dx^\nu $ (invariant under local scale transformations). Since $\partial_t$ is a Killing vector field (in our solutions in this paper), the conserved quantity of the particle is 
\[
P_t  = \left. {\frac{{\partial L}}{{\partial \dot x^t  }}} \right|_\gamma   =  \left. {\phi ^2 \left( x \right)g_{\nu t} \dot x^\nu  } \right|_\gamma  \,,
\]
so
\[
\phi ^2 \left( {x\left( \tau  \right)} \right)g_{tt} \left( {x\left( \tau  \right)} \right)\frac{{dt}}{{d\tau }} = \varepsilon  = constant \, .
\]
Thus, the field $\phi$ experts a force (the right side of (\ref{eq:1116})) on the bodies in addition to the gravitational force. In our solution (\ref{eq:33}), the field $\phi$ is not a pure gauge, thus we can not get rid of this force by any reparameterization. This force works similar to the gravitational force; it is a non-inertial force. Thus, the scalar field in local scale symmetry affects on the bodies as a gravitational force.
\\

If the field $\phi$ is a pure gauge, then by an affine parameterization, the equation (\ref{eq:1116}) can reduce to $ \nabla_\alpha  (d x^\mu/d  \alpha) =0$ which is the usual free fall equation, but the new parameter $\alpha$ is not same the proper time $\tau$. A notable point is that this symmetry does not produce any conserved currents in (\ref{eq:1117}), but the particles are effected by the force in left side of (\ref{eq:1116}).
\\

From $d\tau ^2  =  - \phi ^2 g_{\mu \nu } dx^\mu  dx^\nu $, we get  
\[
g_{\rho \sigma } \left( {x\left( \tau  \right)} \right)\dot x^\rho  \left( \tau  \right)\dot x^\sigma  \left( \tau  \right) =  - \phi ^{ - 2} \left( {x\left( \tau  \right)} \right)\, .
\]
By this, the free fall equation (\ref{eq:1116}) becomes
\[
\nabla _\tau  \dot x^\mu  \left( \tau  \right) =  - 2\dot x^\mu  \left( \tau  \right)\frac{d}{{d\tau }}\ln \phi \left( {x\left( \tau  \right)} \right)\, - \left. {\left[ {\phi ^{ - 2} \left( x \right)g^{\mu \nu } \left( x \right)\partial _\nu  \ln \phi \left( x \right)} \right]} \right|_\gamma  \, .
\]
Using the solutions (\ref{eq:33}), yields (for $x^\mu=r$)
\[
\nabla _\tau  \left( {\frac{{dr}}{{d\tau }}} \right) =  - 2R_\phi  \left( {1 - \frac{r}{{R_\phi  }}} \right)^{ - 1} \left( {\frac{{dr}}{{d\tau }}} \right)^2 \, - \frac{{h^2 R_\phi  }}{{\left( {1 - \kappa } \right)}}\left( {1 - \frac{r}{{R_\phi  }}} \right)^3 \left( {1 - \frac{{2M}}{r}} \right)\, .
\]
In $r< R_\phi$ (for $R_\phi>>R_{bh}=2M$), this force (generated by $\phi$, in the right side) is negative, thus it works to bring the particles into the black hole in $r=0$. 
\\



\section{Solution of black hole of local scale symmetry of $\lambda\ne 0$ } 
In this section, we obtain static solutions for the general case where $\lambda\ne 0$, which produces results that are somewhat more complicated than those found when $\lambda= 0$. The equation $ g_{tt}=0$, now identifies three points, and these points are specified by $M$, $R_ \phi$, $h$, and $ \lambda$, with $ g_{tt}$ reversing its sign at each of them.  As these parameters are independent of one another, it is possible to propose two types of behavior for the field $\phi$: one in which this field is mostly concentrated inside the black hole, and another in which it is mostly concentrated far from the black hole. The first case restricts the field strength parameter $h$ to a certain range. The second case makes the field strength $|\phi|$ decrease without bound, which is what one expects when the black hole's mass is small enough that its gravitational force is unable to gather the scalar field, so that the field $\phi$ becomes concentrated far from the black hole. The solutions we derive resemble those for $\lambda= 0$; the field $\phi$ is largely concentrated close to the surface $r=R_\phi$, and there is no relationship between the horizons $r=2M$ and $r=R_\phi$. In other words, $M$ and $R_\phi$ are independent arbitrary positive parameters that characterize our solutions. Moreover, the solution for $\phi$ is unaffected by the black hole's horizon, since it is a continuously differentiable function at $r=2M$. We ascribe this feature to the scale symmetry, by means of which, as explained below equation (\ref{eq:118}), one can freely modify how the field is distributed throughout space and there are no preferred locations for $\phi$ to concentrate within.
\\

For $\lambda\ne 0$, the Einstein's equation (\ref{eq:63}) has the solution (See Appendix \ref{appendix:b} for details)
\begin{equation}\label{eq:86}
\begin{split}
& ds^2  = - \frac{{a_1 \left( {1 - \frac{{2M}}{r}} \right)}}{{4  R_\phi ^2 \left( {1 - \frac{{2M}}{{R_\phi  }}} \right)^3 }} \left( {r - r_ -  } \right)\left( {r - r_ +  } \right)dt^2  + \frac{{4 R_\phi ^2  \left( {1 - \frac{{2M}}{{R_\phi  }}} \right)^3 }}{{a_1 \left( {1 - \frac{{2M}}{r}} \right) \left( {r - r_ -  } \right)\left( {r - r_ +  } \right)}}dr^2  \\
&\,\qquad+ r^2 d\Omega _2^2 \,,\\
&\phi \left( r \right) = \frac{1}{{h\left(r - R_\phi \right) }}\,.
\end{split}
\end{equation}
Where the horizons $r_{+}$ and $r_{-}$ are given by
\begin{equation}\label{eq:84}
\begin{split}
& r_ \pm   = R_\phi  \frac{{\left( {1 - \frac{{3\lambda }}{{8h^2 }}} \right)\frac{{\left( {2M} \right)^2 }}{{R_\phi^2 }} - \left( {2 - \frac{\lambda }{{8h^2 }}} \right)\frac{{2M}}{R_\phi} + 1}}{{\left( {1 - \frac{{3\lambda }}{{4h^2 }}} \right)\frac{{\left( {2M} \right)^2 }}{{R_\phi^2 }} - \left( {2 - \frac{{3\lambda }}{{4h^2 }}} \right)\frac{{2M}}{R_\phi} + \left( {1 - \frac{\lambda }{{4h^2 }}} \right)}}\\
&\\
 & \pm R_\phi \sqrt {\frac{\lambda }{{h^2 }}} \frac{{\sqrt {\left( {1 - \frac{{3\lambda }}{{16h^2 }}} \right)\frac{{\left( {2M} \right)^4 }}{{4R_\phi^4 }} + \left( { - 1 + \frac{{3\lambda }}{{32h^2 }}} \right)\frac{{\left( {2M} \right)^3 }}{{R_\phi^3 }} + \left( {\frac{3}{2} - \frac{{3\lambda }}{{64h^2 }}} \right)\frac{{\left( {2M} \right)^2 }}{{R_\phi^2 }} + \frac{1}{4}\left( {1 - \frac{{8M}}{R_\phi}} \right)} }}{{\left( {1 - \frac{{3\lambda }}{{4h^2 }}} \right)\frac{{\left( {2M} \right)^2 }}{{R_\phi^2 }} - \left( {2 - \frac{{3\lambda }}{{4h^2 }}} \right)\frac{{2M}}{R_\phi} + \left( {1 - \frac{\lambda }{{4h^2 }}} \right)}}\,,
 \end{split}
\end{equation}
and $a_1$ is given by
\begin{equation}
a_1= \left( {4 - \frac{{3\lambda }}{{h^2 }}} \right)\frac{{\left( {2M} \right)^2 }}{{R_\phi ^2 }} - \left( {4 - \frac{{3\lambda }}{{h^2 }}} \right)\frac{{2M}}{{R_\phi  }} + \left( {4 - \frac{\lambda }{{h^2 }}} \right)\,.
\end{equation}
Unlike the case with $\lambda=0$, the results for $\lambda\ne 0$ do not impose conditions on $\kappa= 2M/R_{\phi} $. As a result, a substantial concentration of the field $\phi$ can be attained inside the black hole (where $R_\phi<2M$), but this requires the condition $a_1 <0$ so that the function $g_{tt}$ remains negative outside the black hole (in $r>r_{+}>r_{-}$). The requirements $R_\phi<2M $ and $a_1 < 0$, together with obtaining real roots in (\ref{eq:84}), entail
\begin{equation}
\frac{4}{{3x^2  - 3x + 1}}\left( {x^2  - x + 1} \right) < \frac{\lambda }{{h^2 }} < \frac{{16}}{{3x^2 }}\left( {x^2  - 2x + 1} \right)\, ,
\end{equation}
for
\begin{equation}
x = \frac{{2M}}{{R_\phi  }} > 2.23 \, .
\end{equation}
By that, the metric (\ref{eq:86}) becomes
 \begin{equation}\label{eq:88}
\begin{split}
\left. {ds^2 } \right|_{R_\phi   < 2M}  =&  - \frac{(-a_1) {\left( {1 - \frac{{2M}}{r}} \right)}}{{4  R_\phi ^2 \left( { \frac{{2M}}{{R_\phi  }} -1  } \right)^3 }} \left( {r - r_ -  } \right)\left( {r - r_ +  } \right)dt^2  + \frac{{4 R_\phi ^2  \left( { \frac{{2M}}{{R_\phi  }}  -1 } \right)^3 }}{{(-a_1)\left( {1 - \frac{{2M}}{r}} \right) \left( {r - r_ -  } \right)\left( {r - r_ +  } \right)}}dr^2  \, \\ 
&\\
&+ r^2 d\Omega _2^2\, , \quad \text{for} \quad a_1<0 \,,
\end{split}
\end{equation}
where $r_{-}$ and $r_{+}$ are the same (\ref{eq:84}). Accordingly, the horizons $r_{-}$, $r_{+}$ and $R_\phi$ (for $R_\phi <2M$) are located inside the black hole. Let us test $ {\lambda }/{{h^2 }} = {{16}}/{{3 }}$ for $x>>1$, which yields
\begin{equation}
r_ +   \approx r_ -   \approx \frac{{R_\phi  }}{3} << 2M\,.
\end{equation}
Thus, in this case, the field $\phi$ mostly resides inside the black hole. We observe that ${\lambda }/{{h^2 }}$ is constrained. Indeed, the parameter ${\lambda }/{{h^2 }}$ originates from the $\phi^4$ term in Einstein's equation (\ref{eq:63}). The $\phi^4$ term yields the parameter ${\lambda }/{{h^4 }}$, though its $1/h^2$ part is dropped when abbreviating with the other $\phi^2$ terms. Hence, ${\lambda }/{{h^2 }}$ serves as the effective coupling constant of the $\phi^4$ term. The restriction on ${\lambda }/{{h^2 }}$ needed to obtain the case $R_\phi  << 2M$ could be a requirement for gathering the field $\phi$ near the black hole; this field becomes concentrated into a smaller volume, so as a result $R_\phi$ decreases, and the field strength inside the black hole (say $|\phi ( 0 ) |=1/(h R_\phi)$) rises quickly, meaning the energy of $\phi^4$ also rises quickly. Thus, keeping the field energy constant demands that ${\lambda }/{{h^2 }}$ be confined to a certain range. This requires the black hole's mass to be large enough to generate a sufficiently strong gravitational force; As we observed (below equation (\ref{eq:67})), the scalar field behaves as though it follows the free fall force.
\\

Another possibility is that the field $\phi$ is concentrated far away from the black hole, $R_\phi>>2M$, which makes the magnitude $|\phi ( r ) |$ small, so $h$ grows large, and this results in $h^2>> \lambda$. Under these conditions, the roots (\ref{eq:84}) approach
\begin{equation}\label{eq:113}
r_ \pm   \approx   \frac{R_\phi}{{\left( {1 - \frac{\lambda }{{4h^2 }}} \right)}}\left( {1 \pm \frac{{\sqrt \lambda  }}{{2h}}} \right)>> 2M\,, \,\,\,\, \text{and}  \,\,\,\,  a_1 \approx 4 \, .
\end{equation}
When $h^2>>\lambda$, it follows that $r_{-}\approx r_{+}$, and $R_\phi$ locates between the two. This situation is similar to what happens when $\lambda=0$. Here, the field $\phi$ stays just a bit within the black hole. Such behavior is expected when the mass $M$ is sufficiently small, meaning the black hole cannot generate enough gravitational force to bring the field inside it.


\section{Introduce thermodynamic phenomenon for the black hole } 
Black hole thermodynamics demonstrates that black holes are not merely the outcome of gravitational collapse into some unknown state of matter; rather, they are thermal objects possessing temperature and entropy, and they follow laws similar to those of classical thermodynamics. As a result, black holes can exist in a thermal equilibrium with their surroundings, with their temperatures as measured by an observer at infinity. As a result, black holes emit radiation, shed energy, and in time evaporate. Many frameworks have been put forward to describe black hole thermodynamics, starting with static black holes, moving on to rotating ones, and then extending to black holes in modified gravity theories \cite{r40}-\cite{r53}. Conversely, quantum fields in imaginary time (with periodicity $\beta = 1/T$) exhibit features resembling those found in statistical physics, having a temperature and a distribution function \cite{r54, r55}. Hence, quantum fields can be coupled to black holes when thermodynamics and statistical physics are taken into account.

In the earlier sections, we derived static solutions to Einstein's equation and arrived at the solutions (\ref{eq:33}), which describe a real scalar field coupled to a black hole when local scale symmetry is taken into account, with $\lambda=0$. We noticed that the behaviour of the scalar field appears as though it obeys the law of force of free fall (equation (\ref{eq:67})). In other words, this is much like saying that the scalar field contains particles, and these particles obey the system's free-fall force. As a result, this static solution merely describes an equilibrium condition arising during the exchange of energy between the gravity of the black hole and the scalar field. This leads us to examine the thermodynamic properties of this system.

The first step is to find the temperature of the medium. Next, we determine the effective mass and the entropy of the black hole. By applying the condition $dS>0$, we confirm that the scalar field moves toward the black hole. This indicates to us that the scalar field accumulates adiabatically in the vicinity of (and within) the black hole. This is consistent with the results below equation (\ref{eq:86}), which point to the possibility that the scalar field is mostly concentrated inside the black hole. As a result, through the continuity of this process, the energy of the scalar field $\phi$ becomes mostly concentrated inside the black hole, so the field $\phi$ may disappear, while the entropy and the effective mass of the black hole grow. Evidence for the accretion of the field $\phi$ into the black hole is provided by the force acting on a freely falling object; this force is directed toward the black hole, so it serves to concentrate the field $\phi$ near the black hole, but this process requires the black hole's mass to be large enough to generate a sufficiently strong gravitational field.
\\

We begin with the metric (\ref{eq:34}):
\begin{equation}\label{eq:65}
ds^2  =  - \frac{1}{{\left( {1 - \kappa } \right)}}\left( {1 - \frac{r}{{R_\phi  }}} \right)^2 \left( {1 - \frac{{2M}}{r}} \right)dt^2  + \frac{{\left( {1 - \kappa } \right)}}{{\left( {1 - \frac{r}{{R_\phi  }}} \right)^2 \left( {1 - \frac{{2M}}{r}} \right)}}dr^2  + r^2 d\Omega _2^2 \,.
\end{equation}
We expand this metric nearby the black hole horizon $r=2M$, by using
\begin{equation}\label{eq:36}
r = 2M\left( {1 + \varepsilon ^2 } \right)\, ,\quad  \text{ for} \quad \varepsilon<<1 \, .
\end{equation}
Taking in consideration only the significant terms of $\varepsilon$, we end with
\begin{equation}\label{eq:37}
\left. {ds^2 } \right|_{r \approx 2M}  =  - \frac{{\left( {R_\phi   - 2M} \right)^2 }}{{\left( {1 - \kappa } \right)R_\phi ^2 }}\varepsilon ^2 dt^2  + \frac{{16M^2 \left( {1 - \kappa } \right)R_\phi ^2 }}{{\left( {R_\phi   - 2M} \right)^2 }}d\varepsilon ^2  + \left( {2M} \right)^2 d\Omega _2^2 \, .
\end{equation}
This metric describes the topology $\mathbb{R}^{1,1} \times S^2$, and it is not a solution of Einstein's equation. It is only an approximate solution nearby the horizon surface $r=2M$. The first two parts of the metric (\ref{eq:37}) can be written in form of 2d Rindler metric in $\mathbb{R}^{1,1} $. The 2d Rindler metric is
\begin{equation}\label{eq:38}
ds^2  = dR^2  - R^2 d\eta ^2 \, .
\end{equation}
There is a horizon at $R = 0$, so these coordinates work for $R > 0$, $\eta =\text{anything}$. The Rindler space (\ref{eq:38}) is related to polar coordinates $ dR^2+R^2 d\theta^2$ by the transformation $\eta = i\theta$. Polar coordinates on $\mathbb{R}^2$ are singular at the origin unless $\theta$ is a periodic coordinate; $\theta\to \theta+2\pi$. Therefore, $\eta $ shall be periodic in the imaginary direction, $\eta \to \eta+2\pi i$. Thus, by approximately writing the metric near horizon black hole in form of Rindler metric, this implies that the time coordinate $t$ has an imaginary periodicity. Let us write (\ref{eq:37}) as
\begin{equation}\label{eq:39}
\left. {ds^2 } \right|_{r \approx 2M}  = \frac{{16M^2 \left( {1 - \kappa } \right)R_\phi ^2 }}{{\left( {R_\phi   - 2M} \right)^2 }}\left( { - \frac{{ \left( {R_\phi   - 2M} \right)^4 }}{{16M^2 R_\phi ^4 \left( {1 - \kappa } \right)^2  }}\varepsilon ^2 dt^2  + d\varepsilon ^2 } \right) + \left( {2M} \right)^2 d\Omega _2^2 \,.
\end{equation}
This implies the imaginary periodicity
\begin{equation}\label{eq:40}
\frac{{\left( {R_\phi   - 2M} \right)^2}}{{4M\left( {1 - \kappa } \right)R_\phi ^2 }}t \sim \frac{{\left( {R_\phi   - 2M} \right)^2}}{{4M\left( {1 - \kappa } \right)R_\phi ^2 }}t + 2\pi i \, ,
\end{equation}
which yields
\begin{equation}\label{eq:41}
 t \sim t + \beta i \, , 
\end{equation}
for
\begin{equation}\label{eq:42}
 \beta  = \frac{{8\pi M\left( {1 - \kappa } \right)R_\phi ^2 }}{{\left( {R_\phi   - 2M} \right)^2}} \, . 
\end{equation}
It is obvious that this periodicity remains invariant under the local scale transformations; ${g_{\mu \nu }}( x ) \to \Omega^2 ( x ){g_{\mu \nu }}( x )$, $\phi( x ) \to \Omega^{-1} ( x ) \phi( x )$. 
\\

According to thermal quantum fields, we obtain the temperature of the black hole:
\begin{equation}\label{eq:43}
T_{bh} = \frac{1}{\beta } = \frac{{\left( {R_\phi   - 2M} \right)^2 }}{{8\pi M\left( {1 - \kappa } \right)R_\phi ^2 }} = \frac{1}{{8\pi M}}\left( {1 - \frac{{2M}}{{R_\phi  }}} \right) \, ,
\end{equation}
where we used $\kappa=R_{bh}/R_\phi =2M / R_\phi <1$. We can make sure that this formula obeys the law
\begin{equation}
 T_{bh} = \frac{{{ - g'_{tt} \left( {R_{bh} } \right)}}}{4\pi }\, .
\end{equation}
Clearly, as $R_\phi$ tends toward infinity, $\kappa=R_{bh}/R_\phi$ approaches zero, and so we recover the standard expression for the Hawking temperature of a black hole with mass $M$. From formula (\ref{eq:43}), and given that $( {1 - {{2M}}/{{R_\phi  }}} )<1$, the scalar field acts to cool the black hole. This may be linked to an increase in the black hole's mass caused by the scalar field's presence. As shown in the scalar field solution, equation (\ref{eq:33}), it does not vanish inside the black hole, which means $\phi(0)=-1/(h R_\phi)$. Conversely, as $R_\phi$ increases, the scalar field becomes largely concentrated away from the black hole, since it is mainly localized near the surface $r=R_\phi$; as a result, its density within the black hole decreases. Hence, the energy inside the black hole likewise decreases. This allows us to write $T=1/(8 \pi M_{eff})$, where
\begin{equation}\label{eq:68}
M_{eff}  = M \left( {1 - \frac{{2M}}{{R_\phi  }}} \right)^{ - 1} \, 
\end{equation}
is the effective mass of the black hole (this agrees with the definition (\ref{eq:119})). Because of the attractive force acting in the region $r<r_0$ (equations (\ref{eq:67}) and (\ref{eq:66})), the black hole traps particles of the field $\phi$; consequently, $R_\phi$ becomes smaller, and therefore the concentration of $\phi$ grows in the vicinity of the black hole (and within it). This is consistent with formula (\ref{eq:68}), in which an increase in $M_{eff}$ corresponds to a decrease in $R_\phi$.
\\

To calculate the entropy of this system, we begin by observing that the employed Lagrangian (in the action (\ref{eq:53})) takes the form $L=\phi^2 R+\cdots$. Thus, we shall apply Wald's entropy formula \cite{r57}. In the case of the black hole, we arrive at the entropy (obtained in Ref. \cite{r57})
\begin{equation}\label{eq:120}
S_{bh} = \frac{{A_H }}{4}\phi ^2 \left( {r_H } \right) = \frac{{4\pi \left( {2M} \right)^2 }}{4}\frac{1}{{h^2 \left( {2M - R_\phi  } \right)^2 }} = \frac{\pi }{{h^2 }}\left( {1 - \frac{{R_\phi  }}{{2M}}} \right)^{ - 2} \, ,
\end{equation}
where $A_H$ denotes the horizon's area computed with respect to the employed metric (\ref{eq:65}), and $\phi(r)$ is the solution (\ref{eq:33}). Clearly, this entropy remains unchanged under the global scale transformation $t\to \alpha t$ and $r\to \alpha r$, which entails $M\to \alpha M$ and $R_\phi  \to \alpha R_\phi $ (with $\alpha$ being a positive constant). Since $R_\phi>2M$, it follows that $S_{bh} $ grows as $R_\phi$ diminishes. The decrease in $R_\phi$ is consistent with the process of $dS>0$ and $dM_{eff}>0$, meaning the surface $r=R_\phi$ moves nearer to the black hole, and the coupled scalar field becomes largely concentrated in the vicinity of (and within) this black hole. This is consistent with our earlier observation (below equation (\ref{eq:67})) that the scalar field behaves as though it follows the free fall force.
\\

Under the local scale transformations, ${g_{\mu \nu }}(x) \to \Omega^2 (x){g_{\mu \nu }}(x)$, the area element $dA(x)$ transforms as $dA( x ) \to  \Omega^2 (x)dA( x )$, thus the entropy (\ref{eq:120}) remains invariant under the local scale transformations; when writing it as an integration over the surface, $S=4^{-1} \int_A \phi ^2 dA$.
\\

Finally, let us examine the behaviour of the metric (\ref{eq:65}) near the surface $r=R_\phi$. Using
\begin{equation}
\frac{r}{{R_\phi  }} = 1 + \varepsilon ^{ - 1} \, ,
\end{equation}
with $\varepsilon \to \infty$, the metric (\ref{eq:65}) approximates to 
\begin{equation}
\left. {ds^2 } \right|_{r \approx R_\phi  ,\, \varepsilon  \approx \infty }  =  - \frac{{\varepsilon ^{ - 2} }}{{\left( {1 - \kappa } \right)}}\left( {1 - \frac{{2M}}{{R_\phi  }}} \right)dt^2  + \frac{{\left( {1 - \kappa } \right)}}{{\varepsilon ^{ - 2} \left( {1 - \frac{{2M}}{{R_\phi  }}} \right)}}R_\phi ^2 \varepsilon ^{ - 4} d\varepsilon ^2  + R_\phi ^2 d\Omega _2^2 \,.
\end{equation}
Using $\kappa=R_{bh}/R_\phi =2M / R_\phi$, yields
\begin{equation}
\left. {ds^2 } \right|_{r \approx R_\phi  ,\, \varepsilon  \approx \infty }  = \varepsilon ^{ - 2} \left( { - dt^2  + R_\phi ^2 d\varepsilon ^2 } \right) + R_\phi ^2 d\Omega _2^2 \, .
\end{equation}
The first component of this metric is a conformally flat Lorentz metric in $\mathbb{R}^{1,1} $ scaled by $\varepsilon ^{ - 2} \to 0$. This constitutes normal behavior, which means it does not take on thermodynamic properties. In other words, in contrast to black holes, the field $\phi$ is not concealed behind a horizon barrier.

\section{Summary}  
We have studied the solution of a static black hole coupled to a real scalar field, taking local scale symmetry into account. Our findings point to two possible outcomes: the first is that the scalar field is mostly concentrated within the black hole. The second possibility is that the scalar field is concentrated at a distance from the black hole. In the first case, the field's strength must stay bounded within a certain range, while in the latter it must decrease without limit. We have derived the black hole's temperature, entropy, and effective mass. We discovered that the rise in entropy matches the accretion of the scalar field onto the black hole, along with an increase in its effective mass.
\\

{\bf Declaration of competing interest}
The authors assert that they do not have any conflict of interest.

{\bf Data availability }
No data was used for the research described in the article.
\\

\appendix
\section{Appendix: Solution of Einstein's equation for $\lambda= 0$}
\label{appendix:a}
We start from the Einstein equations (\ref{eq:98}):
\begin{equation}\label{eq:appa98}
\begin{split}
&\frac{{uv^2  + ruv' - vu}}{{6r^2 v^2 }}\phi ^2  + \frac{1}{6}\frac{u}{v}\phi '^2  - \frac{1}{3}\phi \left( {u\frac{{\phi ''}}{v} - \frac{{uv'}}{{2v^2 }}\phi ' + \frac{{2u\phi '}}{{rv}}} \right) - \frac{\lambda }{8}u\phi ^4  = 0\,,\\
 &\\
&\frac{{ru' - uv + u}}{{6r^2 u}}\phi ^2  + \frac{1}{2}\phi '^2  - \frac{1}{3}\phi \left( { - \frac{{u'}}{{2u}} - \frac{2}{r}} \right)\phi ' + \frac{\lambda }{8}v\phi ^4  = 0\,, \\
 &\\
& \frac{1}{6} \frac{{2r^2 uvu'' - r^2 vu'^2  - r^2 uu'v' + 2ruvu' - 2r\,u^2 v'}}{{4u^2 v^2 }} \phi ^2 \\ 
&\quad - \frac{{r^2 }}{{6v}}\phi '^2  + \frac{{r^2 }}{{3v}}\phi \phi '' + \left( {\frac{{r^2 }}{2}\frac{{u'}}{{uv}} - \frac{{r^2 }}{2}\frac{{v'}}{{v^2 }} + \frac{r}{v}} \right)\frac{{\phi \phi '}}{3} + \frac{\lambda }{8}r^2 \phi ^4  = 0\,.  
 \end{split}
\end{equation}
To render these three equations less complex, we employ the transformation
\begin{equation}\label{eq:99}
u\left( r \right) = r^2 F\left( r \right)\,, \,\,\, v\left( r \right) = r^2 f\left( r \right)\,, \,\,\,  \text{and}\,\,\, \phi \left( r \right) = \frac{{\xi \left( r \right)}}{r}\, ,
\end{equation}
for some independent functions $F( r )$, $f( r )$, and $\xi(r)$. There is no problem with this transformation at the point $r=0$, because this point is a spacetime singularity in black hole solutions, so the solutions are allowed to be undefined there. With these functions, the equations (\ref{eq:appa98}) transform to
\begin{equation}\label{eq:100}
\begin{split}
&\frac{1}{6}Ff\xi '^2  - \frac{1}{3}F\left( {f\xi \xi '' - \frac{1}{2}f'\xi \xi '} \right) + \frac{1}{6}Ff^2 \xi ^2  - \frac{\lambda }{8}Ff^2 \xi ^4  = 0\,,
\\
&\frac{1}{2}Ff\xi '^2  + \frac{1}{6}fF'\xi \xi ' - \frac{1}{6}Ff^2 \xi ^2  + \frac{\lambda }{8}Ff^2 \xi ^4  = 0 \, , 
\\
&  - \frac{1}{{24}}\left( {FF'f' - 2FfF'' + fF'^2 } \right)\xi ^2  - \frac{1}{6}F^2 f\xi '^2  + \frac{1}{6}F\left( {fF' - Ff'} \right)\xi \xi ' \\ 
 & + \frac{1}{3}F^2 f\xi \xi '' + \frac{\lambda }{8}F^2 f^2 \xi ^4  = 0\,. 
\end{split}
\end{equation}
These three equations involve three unknown functions: $F(r)$, $f(r)$, and $\xi(r)$. In this appendix, we solve them for the case $\lambda=0$, where they reduce to
\begin{equation}\label{eq:103}
\begin{split}
&\frac{1}{6}Ff\xi '^2  - \frac{1}{3}F\left( {f\xi \xi '' - \frac{1}{2}f'\xi \xi '} \right) + \frac{1}{6}Ff^2 \xi ^2    = 0\,,
\\
&\frac{1}{2}Ff\xi '^2  + \frac{1}{6}fF'\xi \xi ' - \frac{1}{6}Ff^2 \xi ^2   = 0 \, , 
\\
&  - \frac{1}{{24}}\left( {FF'f' - 2FfF'' + fF'^2 } \right)\xi ^2  - \frac{1}{6}F^2 f\xi '^2  + \frac{1}{6}F\left( {fF' - Ff'} \right)\xi \xi ' \\ 
 & \quad+ \frac{1}{3}F^2 f\xi \xi ''  = 0\,. 
\end{split}
\end{equation}
Let us take $\xi (r) = e^{K(r)} $, where $K(r)$ is some function. Given this, the equations (\ref{eq:103}) turn into
\begin{equation}\label{eq:4}
\begin{split}
&  - \frac{1}{6}F\left( r \right)f\left( r \right)K'^2 \left( r \right) - \frac{1}{3}F\left( r \right)f\left( r \right)K''\left( r \right) + \frac{1}{6}F\left( r \right)f'\left( r \right)K'\left( r \right) + \frac{1}{6}f^2 \left( r \right)F\left( r \right) = 0 \, ,\\ 
 &\\
&\, \, \, \, \,\frac{1}{2}F\left( r \right)f\left( r \right)K'^2 \left( r \right) + \frac{1}{6}f\left( r \right)F'\left( r \right)K'\left( r \right) - \frac{1}{6}f^2 \left( r \right)F\left( r \right) = 0\, , \\  
&\\
&   - \frac{1}{{24}}\left[ {F'\left( r \right)f'\left( r \right)F\left( r \right) - 2F\left( r \right)f\left( r \right)F''\left( r \right) + f\left( r \right)F'^2 \left( r \right)} \right] + \frac{1}{6}F^2 \left( r \right)f\left( r \right)K'^2 \left( r \right)\,\\ 
&  + \frac{1}{3}F^2 \left( r \right)f\left( r \right)K''\left( r \right) + \frac{1}{6}F\left( r \right)f^2 \left( r \right)\left( {\frac{{F\left( r \right)}}{{f\left( r \right)}}} \right)^\prime  K'\left( r \right) = 0\,.
 \end{split}
\end{equation}
These three equations involve three unknown functions: $F(r)$, $f(r)$, and $K(r)$. We start by solving the first two equations, and we find that their solution satisfies the third equation without requiring any additional conditions (in case $\lambda\ne0$, it will be different).
\\

By combining the first two equations of (\ref{eq:4})), we get
\begin{equation}
\frac{1}{3}Ff\left( { - K'^2  + K''} \right) - \frac{1}{6}\left( {Ff} \right)' K' = 0 \, .
\end{equation}
Using the function $\psi  = ( {e^{ - K} } )'$ in this equation, we obtain 
\begin{equation}
\frac{1}{3}Ff\psi' - \frac{1}{6}\left( {Ff} \right)' \psi = 0 \, ,
\end{equation}
which has the general solution
\begin{equation}
\psi ^2  = c^2 Ff \, ,
\end{equation}
for some arbitrary constant $c$. This formula determines the function $\psi$ (so $K$) in terms of the product $Ff$. We get
\begin{equation}\label{eq:6}
\psi  = \left( {e^{ - K} } \right)' = c\sqrt {Ff}  \, .
\end{equation}
This gives the solution for $K$ (and thus $\xi$):
\begin{equation}\label{eq:7}
e^{ - K}  = h + \int {\psi \left( r \right)} dr \Rightarrow \xi \left( r \right) = e^K  = \frac{1}{{h + \int {\psi \left( r \right)} dr}} \, ,
\end{equation}
so,
\begin{equation}\label{eq:8}
K \left( r \right)=  - \ln \left( {h + \int {\psi \left( r \right)} dr} \right) \, , \quad K' \left( r \right)=  -  \frac{\psi \left( r \right)}{ {h + \int {\psi\left( r \right)} dr} } \, ,
\end{equation}
where $h$ is an arbitrary integration constant. 
\\

In these solutions, we have employed the metric (\ref{eq:1}):
\begin{equation}\label{eq:25}
ds^2  =  -u(r) d{t^2} +v(r) d{r^2} +  {r^2}d\Omega _2^2\, ,
\end{equation}
with the transformation (\ref{eq:99}):
\begin{equation}\label{eq:107}
u\left( r \right) = r^2 F\left( r \right)\,, \,\,\, v\left( r \right) = r^2 f\left( r \right)\,, \,\,\,  \text{and}\,\,\, \phi \left( r \right) = \frac{{\xi \left( r \right)}}{r}\, .
\end{equation}
The solutions that are favored for the black holes are the ones satisfying $-g_{tt } g_{rr }=1$. This leads to a volume element $\sqrt{-g} d^4 x$ that is well defined. We therefore impose the condition $-g_{tt } g_{rr }=1$ on our solutions. Using the condition $-g_{tt } g_{rr }=1$ in the metric (\ref{eq:25}) gives $uv=1$. As a result, this yields $Ff=r^{-4}$. Inserting $Ff=r^{-4}$ into the equation $\psi= c \sqrt{Ff}$ (equation \ref{eq:6}) gives
\begin{equation}
\psi(r)=cr^{-2} \, .
\end{equation}
We obtain $K$ from equation (\ref{eq:8}), we get
\begin{equation}\label{eq:27}
\begin{split}
K\left( r \right) = \ln \left( {\frac{r}{{hr - c}}} \right)\,, \,\,\,\,\,\text{and} \,\,\,\,\, \xi \left( r \right) = e^{K \left( r \right) }  = \frac{r}{{hr - c}}\, .
\end{split}
\end{equation}
Hence, the solution of the field $\phi$ is
\begin{equation}\label{eq:108}
 \phi \left( r \right) = \frac{{\xi \left( r \right)}}{r}= \frac{1}{{hr - c}} \, .
\end{equation}
Using $f=r^{-4}/F$, the solutions (\ref{eq:6}) - (\ref{eq:27}) reduce (identify) the first two equations of (\ref{eq:4}) to a single equation: 
\begin{equation}
cr^3 \left( {hr - c} \right)\left( {\frac{{\rm{d}}}{{{\rm{d}}r}}F\left( r \right)} \right) - 3c^2 r^2 F\left( r \right) + h^2 r^2  - 2chr + c^2  = 0\,.
\end{equation}
This equation has a general solution of the form
\begin{equation}\label{eq:26}
F(r) = \frac{{\left( { - hr + c} \right)^2 \left( {Cch^2 r - Cc^2 h + 1} \right)}}{{r^3 hc}} \, ,
\end{equation}
for any arbitrary constant $C$. Luckily, the solutions (\ref{eq:6}) - (\ref{eq:26}) satisfy the third equation of (\ref{eq:4}), while leaving the constant $C$ arbitrary (in case $\lambda\ne0$, it is different). The constant $C$ can be fixed in terms of a black hole's horizon $r=R_{bh}$, at which $F(R_{bh})=0$. This consequently establishes $C$ as
\begin{equation}
C =  - \frac{1}{{ch\left( {hR_{bh}  - c} \right)}}\, .
\end{equation}
By using $c=h R_{\phi}$, where $R_{\phi}$ is an arbitrary constant (which we just use in place of $c$), we obtain
\begin{equation}\label{eq:30}
F\left( r \right) = \frac{{\left( {R_\phi   - r} \right)^2 \left( {r - R_{bh} } \right)}}{{\left( {1 - \frac{{R_{bh} }}{{R_\phi  }}} \right)R_\phi ^2 r^3 }} = \frac{{\left( {1 - \frac{r}{{R_\phi  }}} \right)^2 \left( {1 - \frac{{R_{bh} }}{r}} \right)}}{{\left( {1 - \frac{{R_{bh} }}{{R_\phi  }}} \right)r^2 }} = \frac{{\left( {1 - \frac{r}{{R_\phi  }}} \right)^2 \left( {1 - \frac{{R_{bh} }}{r}} \right)}}{{\left( {1 - \kappa} \right)r^2 }}\,,
\end{equation}
for $\kappa={R_{bh} }/{R_\phi}$. Hence, a black hole horizon is present at $r=R_{bh}$, at which point the function $F(r)$ (and therefore $u(r)$) flips its sign. This solution places a constraint on $R_{\phi}$; it must fulfill $R_{\phi}> R_{bh} $. The solution for the function $f$ follows from $Ff=r^{-4}$, giving
\begin{equation}
f\left( r \right) = \frac{{\left( {1 - \kappa } \right)R_{\phi}^2 }}{{r\left( {R_{\phi} - r} \right)^2 \left( {r - R_{bh} } \right)}} \, .
\end{equation}
Thus, the solutions of the metric functions $u$ and $v$, equation (\ref{eq:107}), are obtained as follows
\begin{equation}
 u\left( r \right)  = \frac{{\left( {R_{\phi} - r} \right)^2 \left( {r - R_{bh} } \right)}}{{\left( {1 - \kappa } \right)R_{\phi}^2 r }} \, , \, \,\,\, \text{and} \, \,\,\, v\left( r \right) = \frac{{\left( {1 - \kappa } \right)R_{\phi}^2 r}}{{\left( {R_{\phi} - r} \right)^2 \left( {r - R_{bh} } \right)}} \, ,
\end{equation}
for $\kappa= R_{bh}/R_{\phi}<1$. In this context, $r=R_{\phi}$ is treated as a scalar field horizon. Thus, the solutions fall into categories defined by two independent arbitrary positive real parameters: one corresponds to the black hole, $R_{bh}$, while the other pertains to the scalar field horizon, $R_{\phi}$.

The solution for the field $\phi$, given by equation (\ref{eq:108}), is
\begin{equation}\label{eq:ab0108}
\phi \left( r \right) = \frac{1}{{hr - c}} = \frac{1}{{h\left( {r - R_{\phi}} \right)}}\, .
\end{equation}
The dimensionless constant $h$ is arbitrary, and this is connected to the symmetry that arises when the field $\phi$ is scaled by a constant value. As for the units, we have $[F]=[r^{-2}]$, $[\phi]=[r^{-1}]$, $[\xi]=0$, $[c]=[r]$ and $[h]=0$.
\\

And the solution of the metric (\ref{eq:25}) is
\begin{equation}\label{eq:ab33}
ds^2  =  - \frac{{\left( {R_\phi   - r} \right)^2 \left( {r - R_{bh} } \right)}}{{\left( {1 - \kappa } \right)R_\phi ^2 r}}dt^2  + \frac{{r\left( {1 - \kappa } \right)R_\phi ^2 }}{{\left( {R_\phi   - r} \right)^2 \left( {r - R_{bh} } \right)}}dr^2  + r^2 d\Omega _2^2 \,.
\end{equation}
It is evident that when $R_\phi $ grows sufficiently large ($R_\phi \to \infty$), the field solution (\ref{eq:ab0108}) approaches zero, while the metric (\ref{eq:ab33}) tends to the Schwarzschild metric (keeping in mind that $\kappa= R_{bh}/R_{\phi} \to 0$). This leads us to identify $R_{bh}$ with $2M$, where $M$ denotes the mass of the black hole. We finally obtain
\begin{equation}\label{eq:ab34}
ds^2  =  - \frac{1}{{\left( {1 - \kappa } \right)}}\left( {1 - \frac{r}{{R_\phi  }}} \right)^2 \left( {1 - \frac{{2M}}{r}} \right)dt^2  + \frac{{\left( {1 - \kappa } \right)}}{{\left( {1 - \frac{r}{{R_\phi  }}} \right)^2 \left( {1 - \frac{{2M}}{r}} \right)}}dr^2  + r^2 d\Omega _2^2 \,.
\end{equation}
The component $g_{tt}$ of this metric vanishes in two points, namely $2M$ and $R_\phi$.

\section{Appendix: Solution of Einstein's equation for $\lambda\ne 0$}
\label{appendix:b}
To obtain the solution of Einstein's equation of the action (\ref{eq:53}) (for $\lambda\ne 0$), we begin with the equations (\ref{eq:100}):
\begin{equation}\label{eq:69}
\begin{split}
&\frac{1}{6}Ff\xi '^2  - \frac{1}{3}Ff\left( {\xi \xi '' - \frac{1}{2}f^{ - 1} f'\xi \xi '} \right) + \frac{1}{6}Ff^2 \xi ^2  - \frac{\lambda }{8}Ff^2 \xi ^4  = 0  \, ,   
\\
&\frac{1}{2}Ff\xi '^2  + \frac{1}{6}fF'\xi \xi ' - \frac{1}{6}Ff^2 \xi ^2  + \frac{\lambda }{8}Ff^2 \xi ^4  = 0\,,
 \\
& - \frac{1}{24}\left({FF'f' - 2FfF'' + fF'^2 }\right)\xi ^2  - \frac{1}{6}F^2 f\xi '^2  + \frac{1}{6}F^2 f\left( {F^{ - 1} F' - f^{ - 1} f'} \right)\xi \xi ' \\
&+ \frac{1}{3}F^2 f\xi \xi '' + \frac{\lambda }{8}f^2 F^2 \xi ^4  = 0 \, .
 \end{split}
\end{equation}
In theses equations, we employed the transformation (equation (\ref{eq:99}))
\begin{equation}\label{eq:110}
u\left( r \right) = r^2 F\left( r \right)\,, \,\,\, v\left( r \right) = r^2 f\left( r \right)\,, \,\,\,  \text{and}\,\,\, \phi \left( r \right) = \frac{{\xi \left( r \right)}}{r}\, ,
\end{equation}
and used the metric (equation (\ref{eq:25}))
\begin{equation}\label{eq:111}
ds^2  = - u (r) d{t^2} +v(r) d{r^2} +  {r^2}d\Omega _2^2\, .
\end{equation}
These three equations contain three unknown functions; $F(r)$, $f(r)$ and $\xi(r)$. By summing the first two equations of (\ref{eq:69}), we obtain
\begin{equation}\label{eq:72}
\frac{2}{3}\left( {Ff} \right)\xi '^2  + \frac{1}{6}\left( {fF} \right)'\xi \xi ' - \frac{1}{3}\left( {Ff} \right)\xi \xi '' = 0\, .
\end{equation}
Using
\begin{equation}
\xi  = e^K \, 
\end{equation}
in equation (\ref{eq:72}), yields
\begin{equation}
\frac{1}{3}\left( {Ff} \right)\left( {K'^2  - K''} \right) + \frac{1}{6}\left( {fF} \right)'K' = 0\, .
\end{equation}
Then using
\begin{equation}\label{eq:74}
\psi  = \left( {e^{ - K} } \right)' \, ,
\end{equation}
we get
\begin{equation}
2\left( {Ff} \right)\psi ' - \left( {fF} \right)'\psi  = 0 \, .
\end{equation}
This has the general solution
\begin{equation}\label{eq:73}
\psi ^2  = c^2 Ff\, ,
\end{equation}
for some arbitrary constant $c$. This formula determines the function $\psi$ (so $K$) in terms of the product $Ff$. We get
\begin{equation}\label{eq:75}
\psi  = \left( {e^{ - K} } \right)' = c\sqrt {Ff}  \, .
\end{equation}
Therefore, we obtain the solution
\begin{equation}\label{eq:76}
 e^{ - K(r)}  = h + \int {\psi \left( r \right) dr}  \Rightarrow \xi (r) = e^{K(r)}  = \frac{1}{{h + \int {\psi \left(r \right) dr} }} \, ,
\end{equation}
where $h$ is an arbitrary constant of integration. As mentioned below equation (\ref{eq:107}), we consider the solution $Ff=r^{-4}$. Using $Ff=r^{-4}$ in the equation $\psi = c\sqrt {Ff}$ (the equation (\ref{eq:75})), yields
\begin{equation}\label{eq:77}
\psi\left( r \right)=cr^{-2} \, .
\end{equation}
With this, the solution (\ref{eq:76}) becomes
\begin{equation}\label{eq:086}
\xi \left( r \right) = \frac{r}{{hr - c}}\,.
\end{equation}
This gives the solution for the field $\phi$ as
\begin{equation}\label{eq:112}
\phi \left( r \right) = \frac{{\xi \left( r \right)}}{r}= \frac{1}{{hr - c}}\, .
\end{equation}
With using $f=r^{-4}/F$, the solutions (\ref{eq:74}) - (\ref{eq:086}) identify the first two equations of (\ref{eq:69}) to a single equation:
\begin{equation}
4c r^3 \left( {hr - c } \right)\left( {\frac{{{d}}}{{{{d}}r}}F\left( r \right)} \right) - 12c^2 r^2 F\left( r \right) + \left(4h^2   - 3\lambda \right) r^2 - 8chr + 4c^2  = 0 \, .
\end{equation}
This equation has the general solution 
\begin{equation}\label{eq:78}
F(r) = A + \frac{B}{r} + \frac{C}{{r^2 }} + \frac{D}{{r^3 }} \, ,
\end{equation}
where the constants $A$, $B$, $C$ and $D$ shall satisfy (when setting the coefficients of powers of $r$ to zeros) the equations:
\begin{equation}\label{eq:79}
12Ac^2  + 4Bch - 4h^2  + 3\lambda= 0\, , \,\, \, \,Bc  + Ch + h = 0\, , \,\,\, \,  Cc  + 3Dh - c  = 0\, .
\end{equation}
Using the solutions (\ref{eq:75}), (\ref{eq:76}), (\ref{eq:77}) and (\ref{eq:78}) in third equation of (\ref{eq:69}), with using $f=r^{-4}/ F$, we obtain
\begin{equation}
\frac{{\left( {12Ac^2  + 8Bch + 4Ch^2  + 3\lambda } \right)r + 4Bc^2  + 8Cch + 12Dh^2 }}{{r^5 \left( { - hr + c} \right)^4 }} = 0\, .
\end{equation}
Thus, we get other two formulas for the constants $A$, $B$, $C$ and $D$:
\begin{equation}\label{eq:80}
12Ac^2  + 8Bch + 4Ch^2  + 3\lambda = 0\, , \, \,\,\, Bc^2  + 2Cch + 3Dh^2  = 0 \, .
\end{equation}
The equations (\ref{eq:79}) and (\ref{eq:80}) have a unique solution (for $h\ne0$):
\begin{equation}\label{eq:81}
B =  - \frac{{C + 1}}{c}h\, , \,\,\, \, D = \frac{{1 - C}}{{3h}}c\, , \, \,\,\, \lambda  =  - 4Ac^2  + \frac{4}{3}Ch^2  + \frac{8}{3}h^2 \, ,
\end{equation}
where the arbitrary constants $C$, $c$ and $h$ are independent each of other. By inserting the solution (\ref{eq:81}) in equation (\ref{eq:78}), and let $F(r)$ vanish in $r = 2M$ (the black hole's horizon), we get a formula for $C$ in terms of the others, and using $c=hR_\phi  $ (just using $R_\phi $ in place of $c$), we get
\begin{equation}\label{eq:83}
{F\left( r \right) = \frac{{\left( {r - 2M} \right)}}{{4R_\phi  \left( {R_\phi   - 2M} \right)^3 r^3 }}\left( {a_1 r^2  + a_2 r + a_3 } \right)} \, ,
\end{equation}
with the coefficients
\begin{equation}\label{eq:82}
\begin{split}
&  a_1= \left( {4 - \frac{{3\lambda }}{{h^2 }}} \right)\frac{{\left( {2M} \right)^2 }}{{R_\phi ^2 }} - \left( {4 - \frac{{3\lambda }}{{h^2 }}} \right)\frac{{2M}}{{R_\phi  }} + \left( {4 - \frac{\lambda }{{h^2 }}} \right)\, ,\\
 &a_2  = R_\phi  \left[ {\left( { - 8 + \frac{{3\lambda }}{{h^2 }}} \right)\frac{{\left( {2M} \right)^2 }}{{R_\phi ^2 }} + \left( {16 - \frac{\lambda }{{h^2 }}} \right)\frac{{2M}}{{R_\phi  }} - 8} \right]\, , \\
&a_3  = R_\phi ^2 \left[ {\left( {4 - \frac{\lambda }{{h^2 }}} \right)\frac{{\left( {2M} \right)^2 }}{{R_\phi ^2 }} + 4\left( {1 - \frac{{4M}}{{R_\phi  }}} \right)} \right]\, .
 \end{split}
\end{equation}
The function (\ref{eq:83}) can be also written as
\begin{equation}
F\left( r \right) = \frac{{a_1\left( {r - 2M} \right)R_\phi}}{{4   \left( {R_\phi   - 2M} \right)^3 r^3 }}\left( {r - r_ -  } \right)\left( {r - r_ +  } \right) \, ,
\end{equation}
for
\begin{equation}\label{eq:appb84}
\begin{split}
& r_ \pm   = R_\phi  \frac{{\left( {1 - \frac{{3\lambda }}{{8h^2 }}} \right)\frac{{\left( {2M} \right)^2 }}{{R_\phi^2 }} - \left( {2 - \frac{\lambda }{{8h^2 }}} \right)\frac{{2M}}{R_\phi} + 1}}{{\left( {1 - \frac{{3\lambda }}{{4h^2 }}} \right)\frac{{\left( {2M} \right)^2 }}{{R_\phi^2 }} - \left( {2 - \frac{{3\lambda }}{{4h^2 }}} \right)\frac{{2M}}{R_\phi} + \left( {1 - \frac{\lambda }{{4h^2 }}} \right)}}\\
&\\
 & \pm R_\phi \sqrt {\frac{\lambda }{{h^2 }}} \frac{{\sqrt {\left( {1 - \frac{{3\lambda }}{{16h^2 }}} \right)\frac{{\left( {2M} \right)^4 }}{{4R_\phi^4 }} + \left( { - 1 + \frac{{3\lambda }}{{32h^2 }}} \right)\frac{{\left( {2M} \right)^3 }}{{R_\phi^3 }} + \left( {\frac{3}{2} - \frac{{3\lambda }}{{64h^2 }}} \right)\frac{{\left( {2M} \right)^2 }}{{R_\phi^2 }} + \frac{1}{4}\left( {1 - \frac{{8M}}{R_\phi}} \right)} }}{{\left( {1 - \frac{{3\lambda }}{{4h^2 }}} \right)\frac{{\left( {2M} \right)^2 }}{{R_\phi^2 }} - \left( {2 - \frac{{3\lambda }}{{4h^2 }}} \right)\frac{{2M}}{R_\phi} + \left( {1 - \frac{\lambda }{{4h^2 }}} \right)}}\,.
 \end{split}
\end{equation}
It is apparent in this solution that the horizons (of $F=0$) depend on the parameters $M$, $h$, $\lambda$ and $R_\phi$. Regarding the units, we have $[F]=[r^{-2}]$, $[\xi]=0$, $[\phi]=[r^{-1}]$, $[c]=[r]$ and $[h]=[\lambda]=0$.
\\

Using the solutions of $F$ and $f=r^{-4} F^{-1}$ in the equations (\ref{eq:110}), the metric (\ref{eq:111}) becomes
 \begin{equation}\label{eq:apb85}
\begin{split}
ds^2  =&  - \frac{{a_1 \left( {1 - \frac{{2M}}{r}} \right)}}{{4  R_\phi ^2 \left( {1 - \frac{{2M}}{{R_\phi  }}} \right)^3 }} \left( {r - r_ -  } \right)\left( {r - r_ +  } \right)dt^2  + \frac{{4 R_\phi ^2  \left( {1 - \frac{{2M}}{{R_\phi  }}} \right)^3 }}{{a_1 \left( {1 - \frac{{2M}}{r}} \right) \left( {r - r_ -  } \right)\left( {r - r_ +  } \right)}}dr^2  \\
&+ r^2 d\Omega _2^2 \,.
\end{split}
\end{equation}
While the field $\phi$, equation (\ref{eq:112}), is
\begin{equation}\label{eq:apb86}
\phi \left( r \right) = \frac{1}{{h\left(r - R_\phi \right) }}\,.
\end{equation}
Unlike the case $\lambda=0$, in $\lambda\ne0$, the metric component $g_{tt}$ vanishes in three points. And there are no conditions on $\kappa=2M/R_\phi$.

\end{document}